\documentclass[conference]{IEEEtran}
\IEEEoverridecommandlockouts

\def\BibTeX{{\rm B\kern-.05em{\sc i\kern-.025em b}\kern-.08em
    T\kern-.1667em\lower.7ex\hbox{E}\kern-.125emX}}

\usepackage[english]{babel}
\usepackage[utf8]{inputenc}

\usepackage{xspace}
\xspaceaddexceptions{\%}
\xspaceremoveexception{-}

\usepackage{csquotes}

\usepackage{enumitem}
\setitemize{noitemsep,topsep=0pt,parsep=0pt,partopsep=0pt,leftmargin=15pt}
\setenumerate{noitemsep,topsep=0pt,parsep=0pt,partopsep=0pt,leftmargin=20pt}

\newcommand{\bug}{bug\xspace}
\newcommand{\bugs}{bugs\xspace}

\newcommand{\Bugs}{Bugs\xspace}
\newcommand{\buggy}{buggy\xspace}

\newcommand{\Q}{Quantum\xspace}
\newcommand{\q}{quantum\xspace}

\newcommand{\mytitle}{QBugs: A Collection of Reproducible \Bugs in Quantum
Algorithms and a Supporting Infrastructure to Enable Controlled Quantum Software
Testing and Debugging Experiments}

\usepackage[numbers]{natbib}
\usepackage{url}
\usepackage{xcolor}
\definecolor{bluecolour}{rgb}{0,0.2,0.6}
\definecolor{greencolour}{rgb}{0,.5,0}
\definecolor{marooncolour}{cmyk}{0, 0.87, 0.68, 0.32}
\usepackage{hyperref}
\hypersetup{
  pdftitle={\mytitle},
  pdfauthor={Jos{\'e} Campos, Andr\'{e} Souto},
  colorlinks=true,
  bookmarksopen,
  bookmarksnumbered,
  filecolor=greencolour,
  urlcolor=bluecolour,
  linkcolor=marooncolour,
  citecolor=marooncolour,
}

\usepackage[hang,flushmargin]{footmisc}

\usepackage{balance}

\begin{document}

%% ACM
% \title[QBugs]{\mytitle}
%% IEEE
\title{\mytitle}


%% ACM
% \author{Jos\'{e} Campos{\dag}, Andr\'{e} Souto{\dag\S}}
% \affiliation{%
%   \institution{%
%     {\dag}~LASIGE, Faculdade de Ciências, Universidade de Lisboa, Lisboa, Portugal \\
%     {\S}~Instituto de Telecomunicações, Lisboa, Portugal
%   }
%   \city{}
%   \country{}
% }
% \email{jcmcampos@fc.ul.pt, ansouto@fc.ul.pt}

%% IEEE
\author{%
  \IEEEauthorblockN{Jos\'{e} Campos{\dag}, Andr\'{e} Souto{\dag\S}}
  \IEEEauthorblockA{%
    {\dag}LASIGE, Faculdade de Ciências, Universidade de Lisboa, Portugal, and {\S}Instituto de Telecomunicações, Lisboa, Portugal \\
    % {\S}~Instituto de Telecomunicações, Lisboa, Portugal\\
  jcmcampos@fc.ul.pt, ansouto@fc.ul.pt
}}

%% IEEE
\maketitle


\begin{abstract}
  Reproducibility and comparability of empirical results are at the core tenet
  of the scientific method in any scientific field.  To ease reproducibility of
  empirical studies, several benchmarks in software engineering research, such
  as Defects4J, have been developed and widely used.  For \q software
  engineering research, however, no benchmark has been established yet.  In this
  position paper, we propose a new benchmark---named QBugs---which will provide
  \emph{experimental subjects} and an experimental infrastructure to ease the
  evaluation of new research and the reproducibility of previously published
  results on \q software engineering.
  %
  % It allows we 
  % \Q computing is an emerging research area that will, likely, enables us to
  % solve several computational problems more efficiently than it is possible on
  % classic computers.
  % Michael A Nielsen and Isaac Chuang. 2002. Quantum Computation and Quantum Information
  % John Preskill. 2018. Quantum Computing in the NISQ era and beyond. Quantum 2 (2018) https://doi.org/10.22331/q-2018-08-06-79
  % The Quantum Software Lifecycle, https://www.iaas.uni-stuttgart.de/publications/Weder2020_QuantumSoftwareLifecycle.pdf
\end{abstract}


%% ACM
% \keywords{Software engineering, \Q software testing, Software \bugs}
%% IEEE

%% ACM
% \maketitle

\section{Problem Statement}

Despite an endless number of fields where \q computing could overpass
traditional computing (e.g., factoring numbers~\cite{shor}, perform unstructured
search~\cite{grover}, optimization algorithms~\cite{farhi2019quantum}), solve
linear equations~\cite{hhl}), \q software---as \emph{classic} software---needs
to be specified, developed, and tested by human developers.  On one hand, these
tasks have been well studied by researchers and innumerous approaches have been
proposed and evaluated in \emph{classic} software.  On the other hand, in the \q
world, the execution of these tasks implies novel research in several directions
as well as the development of new approaches and prototypes~\cite{zhao2020}.  \Q
software engineering research is still in its early days, as it is the empirical
evaluation of novel approaches in \q computing.  To begin with
\begin{displayquote}
  \emph{How would one evaluate their own novel \q software engineering research
  or, reproduce others research on \q software engineering?}
\end{displayquote}

In its classic counterpart, research ideas and prototypes have been evaluated
and reproduced through \emph{controlled} empirical experiments, which usually
require:
\begin{itemize}
  \item \emph{Experimental subjects}: software artifacts in the form of source
  code, test cases, documentation, \bug reports, and commit history.
  \item \emph{Experimental infrastructure}: configuration files, scripts, and
  tools.
\end{itemize}
To ease the burden, to some extent, of (1) finding representative and diverse
\emph{experimental subjects}, (2) (re-)developing an \emph{experimental
infrastructure}, and (3) reproducing others research, several databases of
software artifacts have been
developed~\cite{sir,d4j,BugsDotJar,BugsJS,Codeflaws}.

Empirical experiments in \q software engineering are still at its
infancy~\cite{zhao2020} and, to the best of our knowledge, no database of \q
software artifacts have been presented.  The lack of reusable artifacts and
datasets on \q software engineering may hold the research community back and
reduce confidence in empirical results.  In fact, we (community) are already
witnessing the same \emph{reproducibility} issue we have been trying to address
in its classic counterpart.  Recently, \citet{liumutation} empirical evaluated
the effectiveness and performance of their novel compiler-level optimization
approach of \q programs, on a benchmark that is no longer
available\footnote{\url{https://sites.google.com/site/qbenchmarks}}
and (to the best of our knowledge) cannot be re-created. Thus, how are others
supposed to reproduce the study?

Lack of reproducibility goes beyond missing artifacts.  Setting up a local
environment similar to the environment other researchers used in the past
to run a particular experimental analysis is, unfortunately, the most
frustrating experience of being a researcher in software engineering.
Inconsistent program versions, long installations steps that failed through the
process, and obscure setup instructions are among the most common problems faced
by researchers when they try to reproduce an experimental analysis.  Virtual
machines and docker containers only partially have solved this problem.
Although both allow one to easily re-run an experimental setup on another
computer, they do not provide a way to know how all the different modules of an
experimental setup are connected, how to augment the defined setup, or how to
run a pre-defined experiment with different inputs, etc.

Reproducibility (or the lack of it) of others' research is an increasing concern
in software engineering~\cite{rep}, as shown by papers with conflicting results
and by the recent establishment of artifact evaluation committees on top-tier
conferences (e.g.,
POPL\footnote{\url{https://popl21.sigplan.org/track/popl-2021-artifact-evaluation}},
ICSE\footnote{\url{https://conf.researchr.org/track/icse-2021/icse-2021-artifact-evaluation}}).
We foresee that reproducibility will also be an issue in \q software engineering
research if, we (as community), do not establish and provide \emph{experimental
subjects and infrastructures}.

% a result obtained by an experiment or observational study should be achieved
% again with a high degree of agreement when the study is replicated with the same
% methodology by different researchers.  To reproduce an empirical study one would
% need to use the same \emph{experimental subjects} and the same
% \emph{experimental infrastructure} as the original study.

Thus, in this position paper, we propose the development of a novel framework
named QBugs for \q software testing and debugging research which aims to
provide:
\begin{enumerate}
  \item A catalog of open-source \q algorithms.
  \item A catalog of reproducible \bugs in \q algorithms.
  \item Supporting infrastructure to enable controlled empirical experiments.
\end{enumerate}
In the following we discuss the challenges to develop QBugs and research
opportunities that could be built on top of QBugs.

% In order to provide such support, QBugs will require and be organized in three
% main modules 
% %
% \subsubsection*{\Bug mining}
% %
% This module will be responsible for ...
% 
% \subsubsection*{\Bug reproducibility}
% %
% This module will be responsible for ...
% 
% \subsubsection*{Test execution}
% %
% This module will be responsible for ...

\section{Challenges}
\noindent
This section describes the %three
main challenges to develop QBugs.

\subsection{\Q Programming Languages}

First main challenge a framework such as QBugs would have to address is the
support for different \q programming languages~\cite{qlanguages}, e.g.,
Q\#~\cite{qsharp}, OpenQASM~\cite{openqasm}, Cirq~\cite{cirq},
Quipper~\cite{quipper}, and Scaffold~\cite{scaffold}.  In order to build the
most diverse collection of reproducible \bugs in quantum algorithms, we will
provide support for the most adopted and supported \q programming languages.

\subsection{Open-Source Implementations of \Q Algorithms}

Coding \q algorithms is very difficult and requires experts to do it.  As
opposed to its classic counterpart, finding open-source projects that implement
\q algorithms is a rather difficult task but crucial for the success of QBugs.
With no open-source implementations of \q algorithms there are no \bugs to mine
from, and therefore no \emph{experimental subjects or infrastructure}.
%
% Stephen Jordan created an online comprehensive catalog of 64 \q
% algorithms\footnote{Comprehensive catalog of \q algorithms
% \url{https://quantumalgorithmzoo.org}, last updated on November 18, 2019,
% visited on January 21, 2021.}.  However, only a few of them have a
% open-source implementation available.
%

To address this challenge and build a prototype of QBugs, we will rely on the
open-source implementations of \q algorithms that live in the \q framework
repositories.  To the best of our knowledge, there are at least three sources of
mature implementations of \q algorithms / programs that we could consider.
\begin{itemize}
  \item ProjectQ's framework repository\footnote{\url{https://github.com/ProjectQ-Framework/ProjectQ/tree/c15f3b2/examples}}
  includes the implementation of 12 \q algorithms.

  \item Qiskit-Aqua's repository\footnote{\url{https://github.com/Qiskit/qiskit-aqua/tree/a8ab494/qiskit/aqua/algorithms}}
  includes the implementation of 29 \q algorithms developed in Qiskit, including
  the successful and well known Shor~\cite{shor}, Grover~\cite{grover}, and
  HHL~\cite{hhl} algorithms.

  \item
  Repository\footnote{\url{https://github.com/oreilly-qc/oreilly-qc.github.io/tree/1a4c2cc/samples}}
  of the book ``Programming Quantum Computers'' from O'Reilly~\cite{qoreilly}
  includes, overall, 166 \q exercises and correspondent solutions (some implementing \q
  algorithms) written in 6 different \q frameworks: 10 algorithms written in
  Cirq, 4 in DWave, 29 in OpenQASM, 54 in QCEngine, 40 in Q\#, and 29 in Qiskit.
\end{itemize}

\subsection{\Bugs Mining}

The second challenge is related to the \bug mining procedure.  Given the
implementation of a \q algorithm, how could we automatically identify which
\bugs have been reported (thus real \bugs), and what was the fix?

To address this challenge, we will adopt the same procedure \citet{d4j} and
\citet{BugsJS} used to create the Defects4J database and the BugsJS database,
respectively.  In detail, we will first build a procedure to automatically map
each \bug report, that was labeled as ``bug'' or ``issue'' in the issue platform
(e.g., Github issues), to a commit message in order to find the commit that
fixed the \bug.  Then, we will identify the \bug commit as being the commit
right before the \bug fixing commit and add that \bug to our catalog.  Others
have started studying which \bugs may raise from some \q algorithms
implementations~\cite{huang2018qdb,huang2020}.  We, on the other hand, will
\emph{automatically} extract \bugs from the repository's history.

A preliminary investigation of this challenge reveals, for example, that 93
issues were labeled as ``type: bug'' and closed in the Qiskit-Aqua's repository.
For example, issue
\#928\footnote{\url{https://github.com/Qiskit/qiskit-aqua/issues/928}} reported
a \bug in the implementation of the QAOA algorithm~\cite{farhi2014quantum},
which is then addressed in commit
5695f9f\footnote{\url{https://github.com/Qiskit/qiskit-aqua/commit/5695f9f}}
with the message: ``Fix \#928''.  Note that as repositories such as ProjectQ's
framework repository, Qiskit-Aqua's repository, among others, also include the
source code of the \q language, framework, or simulator, some \bug reports
marked as ``bug'' might not be related to a \bug in any \q algorithm, but
related to a \bug in the \q framework.  For example, issue
\#1324\footnote{\url{https://github.com/Qiskit/qiskit-aqua/issues/1324}} (also
labeled as ``bug'') was related to a \bug in the \q framework.  The proposed
fix\footnote{\url{https://github.com/Qiskit/qiskit-aqua/pull/1340/files}} only
modified files under \texttt{qiskit/aqua/operators/} and the source code of \q
algorithms lives under \texttt{qiskit/aqua/algorithms/}.  QBugs' \bug mining
module will have to be able to distinguish between framework's \bugs and \q
algorithm's \bugs in order to create an accurate catalog.

% In software engineering research, the use of real \bugs has yielded realistic
% experiments and has provided stronger empirical evidence~\cite{fl} than
% artificial \bugs.  For such reason, the second challenge we will face while
% developing QBugs is the mining of \bugs.

\subsection{\Bugs Reproducibility}

Another challenge in QBugs is the reproducibility of \bugs.  \Q programs are not
by design deterministic.  Thus, given a \buggy version of a \q algorithm, is it
possible to automatically reproduce/trigger the \buggy behavior in a
deterministic way?

According to a recent study conducted by~\citet{fingerhuth2018}, automated tests
of \q algorithms appear to be very popular: 23 open-source \q projects out of 26
have tests in place.  Furthermore, \citet{fingerhuth2018} also concluded that
the ratio of code exercise by the tests is slightly above the industry-expected
standard (87\% vs. 85\%).  Research on statistical test oracles for \q
algorithms have been proposed~\cite{huang2020} which may further leverage the
use of software tests in \q computing.

One possible avenue to address this challenge is, therefore, to use the
project's tests to run a \buggy version of a \q algorithm and to assess whether
a \bug is reproducible. For \q projects with no tests in place, e.g., the
exercises in the book ``Programming Quantum Computers'' from
O'Reilly~\cite{qoreilly}, another methodology would have to be defined (e.g.,
runtime oracles~\cite{li2020,liu2020,liusystematic}).

\section{Opportunities}

This section describes the several research and teaching opportunities that
could be developed on top of QBugs.

\subsection{Research opportunities}

Research areas that will immediately benefit from QBugs include: \q software
testing (e.g., regression testing, mutation testing, automatic test generation)
and \q debugging (e.g., fault localization), program comprehension, automated
program repair, software evolution, mining software repositories, and machine
learning~\cite{machQ} in \q software engineering.

For instance, once a prototype of QBugs is available, it would be interesting to
investigate how are \bugs introduced in \q algorithms, type of \q \bugs, and how
are \q \bugs fixed.  This could leverage research on novel static/dynamic tools
tailored to identify \q \bugs on \q programs.  The catalog of real \bugs will
also allow researchers to develop and evaluate the effectiveness of, e.g., fault
localization techniques at identifying the components that are more likely to be
faulty, or the effectiveness of test generators at generating tests that trigger
the faulty behavior.

In summary, QBugs will:
\begin{itemize}
  \item Enable researchers to perform realistic experiments on real \q software
  and on real \q \bugs.
  \item Allow researchers to focus on research ideas by freeing them from the
  task of (re-)developing an experimental infrastructure.
  \item Foster reproducibility and comparability of empirical studies by
  providing reusable artifacts and datasets.
\end{itemize}

\subsection{Teaching opportunities}

We also foresee that QBugs will be useful for software testing and \q classes at
undergraduate or graduate level.  Instructors would have a large pool of
artifacts from which they can select the best ones that might support the
learning goals of their software testing or quantum-related course.  Students
will be able to explore how \q algorithms have been developed and tested,
experiment previously proposed \q software engineering research, reproduce
previously published results, and try new ideas/prototypes.

\section{Related Work}

For the \q software engineering's classical counterpart, several catalogs of
real and artificial software faults~\cite{sir,d4j,BugsDotJar,BugsJS,Codeflaws}
for different programming languages have been proposed and widely accepted by
the research community.  For example, the Software-artifact Infrastructure
Repository (SIR)~\cite{sir} is considered the first attempt to provide a
database of \bugs to enable reproducibility in software testing research.  SIR
provides 81 programs written in Java, C, C++, and C\# and most of the faults are
hand-seeded or obtained from program mutation.  Defects4J~\cite{d4j} is another
well known dataset that contains 835 real \bugs from 17 Java projects, and it
has been used for benchmarking and comparing (1) automatic test generation
approaches~\cite{genTestRealFaults}, (2) fault localization
techniques~\cite{fl}, (3) automatic program repair
techniques~\cite{martinez2015automatic}; and it also been used to study the
properties of the \bugs and their characteristics~\cite{bugDissection}.  Besides
providing real \bugs, Defects4J also provides uniform access to the \bugs
through its own API by abstracting away the version control system (e.g., git)
and build system.

We, on the other hand, aim to (1) gather a significant number of real \bugs that
have been reported to open-source \q algorithms and therefore are quantum-based, and
(2) integrate into our framework previously proposed tools / techniques on \q
software testing and debugging to ease future evaluations and comparisons (e.g.,
the mutation testing tool for quantum computing MTQC~\cite{mtqc}).

\section{Conclusion}

A core tenet of the scientific method is reproducibility of experiments.  To
reproduce an empirical study in, e.g., software testing, one would require to
use the same \emph{experimental subjects}, i.e., same software projects and/or
same catalog of software \bugs, and the same \emph{experimental infrastructure},
e.g., scripts and tools.  Although the former and the latter have been
investigated in detail in software engineering, there is not yet a well
defined benchmark to ease reproducibility in \q computing.  Given the lack of
reproducibility in software engineering~\cite{rep}, we foresee that such concern
will also be valid in \q computing in the near future.  Thus, in this position
paper, we propose a framework which will provide a catalog of \q \bugs and an
infrastructure to conduct empirical experiments.  We envisage that our framework
will have a large potential for future research in the field, in particular
reproducibility and comparability of empirical studies.  For instance,
researchers will be able to easily evaluate and compare the performance of
different techniques under a common setup.

%% ACM
% \begin{acks}
% \end{acks}
%% IEEE
\section*{Acknowledgment}
This work was supported
by the Instituto de Telecomunicações (IT) Research Unit,
  ref.\ UIDB/EEA/50008/2020, granted by FCT/MCTES;
by the %Portuguese Government through
Foundation for Science and Technology (FCT),
  project QuantumMining ref.\ POCI-01-0145-FEDER-031826 and
  project QuantumPrime ref.\ PTDC/EEI-TEL/8017/2020;
by the FEDER through the %Competitiveness and Internationalization Operational Programme (COMPETE 2020)
COMPETE 2020 programme and by the Regional Operational Program of Lisboa,
  project Predict ref.\ PTDC/CCI-CIF/29877/2017;
by the EU,
  project ref.\ UIDB/50008/2020-UIDP/50008/2020 (action QuRUNNER);
and by the LASIGE Research Unit, ref.\ UIDB/00408/2020 and ref.\ UIDP/00408/2020.

%% ACM
% \bibliographystyle{ACM-Reference-Format}
% \bibliography{paper}
%% IEEE
\balance
\bibliographystyle{IEEEtranN}
\bibliography{paper}

\end{document}